\documentclass[letterpaper,twocolumn,superscriptaddress,floatfix,iopjournal]{revtex4-2}

\usepackage{inputenc}
\usepackage{bm}
\usepackage{multirow,amssymb,amsbsy,amsmath}
\usepackage{graphicx}
\usepackage{float}
\usepackage{verbatim}
\usepackage{color}
\makeatletter
\usepackage{pifont}
\usepackage{siunitx}
\usepackage{upgreek}
\usepackage{color}
\usepackage{indentfirst}
\usepackage{soul}
\usepackage{braket}
\usepackage[colorlinks=true,linkcolor=blue,citecolor=blue,urlcolor=blue]{hyperref}

\newcommand{\Rmnum}[1]{\expandafter\@slowromancap\romannumeral #1@}
\usepackage{etoolbox}
\usepackage{ragged2e}
\makeatletter
\apptocmd{\@caption}{\justifying}{}{}
\makeatother

\makeatletter
\renewcommand{\section}{\@startsection{section}{1}{0mm}
	{-\baselineskip}{0.5\baselineskip}{\bf\leftline}}
\makeatother

\begin{document}
\title{Decoherence-protected entangling gates in a silicon carbide quantum node}

\author{Shuo Ren}
\thanks{These authors contribute equally to this work\label{Contribute}}
\affiliation{Laboratory of Quantum Information, University of Science and Technology of China, Hefei, Anhui 230026, China}
\affiliation{Anhui Province Key Laboratory of Quantum Network, University of Science and Technology of China, Hefei, Anhui 230026, China}
\affiliation{CAS Center For Excellence in Quantum Information and Quantum Physics, University of Science and Technology of China, Hefei, Anhui 230026, China}
\affiliation{Hefei National Laboratory, University of Science and Technology of China, Hefei, Anhui 230088, China}

\author{Rui-Jian Liang}
\thanks{These authors contribute equally to this work\label{Contribute}}
\affiliation{Laboratory of Quantum Information, University of Science and Technology of China, Hefei, Anhui 230026, China}
\affiliation{Anhui Province Key Laboratory of Quantum Network, University of Science and Technology of China, Hefei, Anhui 230026, China}
\affiliation{CAS Center For Excellence in Quantum Information and Quantum Physics, University of Science and Technology of China, Hefei, Anhui 230026, China}

\author{Zhen-Xuan He}
\affiliation{Laboratory of Quantum Information, University of Science and Technology of China, Hefei, Anhui 230026, China}
\affiliation{Anhui Province Key Laboratory of Quantum Network, University of Science and Technology of China, Hefei, Anhui 230026, China}
\affiliation{CAS Center For Excellence in Quantum Information and Quantum Physics, University of Science and Technology of China, Hefei, Anhui 230026, China}
\affiliation{Hefei National Laboratory, University of Science and Technology of China, Hefei, Anhui 230088, China}
   
\author{Ji-Yang Zhou}
\affiliation{Laboratory of Quantum Information, University of Science and Technology of China, Hefei, Anhui 230026, China}
\affiliation{Anhui Province Key Laboratory of Quantum Network, University of Science and Technology of China, Hefei, Anhui 230026, China}
\affiliation{CAS Center For Excellence in Quantum Information and Quantum Physics, University of Science and Technology of China, Hefei, Anhui 230026, China}

\author{Wu-Xi Lin}
\affiliation{Laboratory of Quantum Information, University of Science and Technology of China, Hefei, Anhui 230026, China}
\affiliation{Anhui Province Key Laboratory of Quantum Network, University of Science and Technology of China, Hefei, Anhui 230026, China}
\affiliation{CAS Center For Excellence in Quantum Information and Quantum Physics, University of Science and Technology of China, Hefei, Anhui 230026, China}
\affiliation{Hefei National Laboratory, University of Science and Technology of China, Hefei, Anhui 230088, China}	
   
\author{Zhi-He Hao}
\affiliation{Laboratory of Quantum Information, University of Science and Technology of China, Hefei, Anhui 230026, China}
\affiliation{Anhui Province Key Laboratory of Quantum Network, University of Science and Technology of China, Hefei, Anhui 230026, China}
\affiliation{CAS Center For Excellence in Quantum Information and Quantum Physics,
University of Science and Technology of China, Hefei, Anhui 230026, China}

\author{Bing Chen}
\affiliation{School of Physics, Hefei University of Technology, Hefei, Anhui 230009, China}

\author{Tao Tu}
\altaffiliation{Email: tutao@ustc.edu.cn}
\affiliation{Laboratory of Quantum Information, University of Science and Technology of China, Hefei, Anhui 230026, China}
\affiliation{Anhui Province Key Laboratory of Quantum Network, University of Science and Technology of China, Hefei, Anhui 230026, China}
\affiliation{CAS Center For Excellence in Quantum Information and Quantum Physics, University of Science and Technology of China, Hefei, Anhui 230026, China}
\affiliation{Hefei National Laboratory, University of Science and Technology of China, Hefei, Anhui 230088, China}

\author{Jin-Shi Xu}
\altaffiliation{Email: jsxu@ustc.edu.cn}
\affiliation{Laboratory of Quantum Information, University of Science and Technology of China, Hefei, Anhui 230026, China}
\affiliation{Anhui Province Key Laboratory of Quantum Network, University of Science and Technology of China, Hefei, Anhui 230026, China}
\affiliation{CAS Center For Excellence in Quantum Information and Quantum Physics, University of Science and Technology of China, Hefei, Anhui 230026, China}
\affiliation{Hefei National Laboratory, University of Science and Technology of China, Hefei, Anhui 230088, China}

\author{Chuan-Feng Li}
\altaffiliation{Email: cfli@ustc.edu.cn}
\affiliation{Laboratory of Quantum Information, University of Science and Technology of China, Hefei, Anhui 230026, China}
\affiliation{Anhui Province Key Laboratory of Quantum Network, University of Science and Technology of China, Hefei, Anhui 230026, China}
\affiliation{CAS Center For Excellence in Quantum Information and Quantum Physics, University of Science and Technology of China, Hefei, Anhui 230026, China}
\affiliation{Hefei National Laboratory, University of Science and Technology of China, Hefei, Anhui 230088, China}
 
\author{Guang-Can Guo}
\affiliation{Laboratory of Quantum Information, University of Science and Technology of China, Hefei, Anhui 230026, China}
\affiliation{Anhui Province Key Laboratory of Quantum Network, University of Science and Technology of China, Hefei, Anhui 230026, China}
\affiliation{CAS Center For Excellence in Quantum Information and Quantum Physics, University of Science and Technology of China, Hefei, Anhui 230026, China}
\affiliation{Hefei National Laboratory, University of Science and Technology of China, Hefei, Anhui 230088, China}

\begin{abstract}

Solid-state color centers are promising candidates for nodes in quantum network architectures. However, realizing scalable and fully functional quantum nodes—comprising both processor and memory qubits with high-fidelity universal gate operations—remains a central challenge in this field. Here, we demonstrate a fully functional quantum node in silicon carbide (SiC), where electron spins act as quantum processors and nuclear spins serve as quantum memory. Specifically, we design a pulse sequence that combines dynamical decoupling with hyperfine interactions to realize decoherence-protected universal gate operations between the processor and memory qubits. Leveraging this gate, we deterministically prepare entangled states within the quantum node, achieving a fidelity of $90 \pm 3\%$, which exceeds the fault-tolerance threshold of certain quantum network architectures. These results open a pathway toward scalable, fully functional quantum nodes based on SiC.

\end{abstract}
	
  \maketitle
  
  \date{\today}
  
\vspace{10pt}\noindent\textsf{\textbf{\large  1. Introduction}}

Quantum network architectures are essential for a wide range of quantum technology applications, including long-distance quantum communication, distributed quantum computing, and quantum sensing networks \cite{Kimble2008,Sangouard2011,Northup2014,Wehner2018}. Such networks typically consist of stationary quantum nodes interconnected by flying photonic channels, with various physical platforms proposed as candidates for the realization of quantum nodes \cite{Awschalom2018,Rodgers2018,Aharonovich2016,Reiserer2015,Duan2010}.

In recent years, color centers in silicon carbide (SiC) have emerged as a highly promising solid-state platform for quantum networks and have attracted significant attention \cite{Koehl2011,Widmann2015,Christle2017,Anderson2019,Mu2020,Wang2020,Hesselmeier2024siv,Xu2015SiCnetwork,Christle2015,Seo2016,Li2022,Nagy2019,Bourassa2020,Babin2022,Fang2024,Falk2015,Crook2020,Klimov2015,Lai2024,Hesselmeier2024,Hu2024,Zhou2023,Son2022}. SiC hosts multiple types of color centers with long spin coherence times and exhibits near-infrared optical transitions, making it compatible with photons in the telecommunication band. Furthermore, as a wide-bandgap semiconductor, SiC benefits from mature wafer-scale microfabrication technologies, offering unique advantages in device integration, scalability, and compatibility with existing semiconductor infrastructure.

Despite these notable breakthroughs, current demonstrations are typically limited to single-functional entangled operations on spin color centers per node~\cite{Klimov2015,Bourassa2020,Hu2024},  thereby constraining both the versatility of quantum nodes and the scalability of quantum networks. A fully functional quantum node requires two essential components: a processor and a memory, along with universal gate operations capable of creating and manipulating entanglement between them. Achieving such functionality in SiC remains an ambitious yet highly challenging goal. On one hand, the complex spin energy level structure and decoherence mechanisms inherent to the SiC system pose significant difficulties, with no clear solution currently established. On the other hand, realizing high-fidelity entangled-state manipulation that meets the stringent fault-tolerance thresholds of quantum networks adds further technical challenges in the solid-state environment.

Coherent control of coupled electron–nuclear spin systems is well established in diamond nitrogen-vacancy (NV) centers, where electron-spin dynamical decoupling (DD) combined with radio-frequency (RF) control of nuclear spins enables long-lived quantum memories and universal electron–nuclear gate operations~\cite{Smeltzer2009RobustControl,deLange2010,Ryan2010,Taminiau2012,vanDerSar2012,Bradley2019}.
Extending this powerful control paradigm to SiC, however, remains nontrivial. Previous studies in SiC have largely focused on weakly coupled nuclear spins and relied predominantly on electron-spin dynamical decoupling, without explicit RF driving or phase-calibrated control of the nuclear degrees of freedom~\cite{Babin2022,Bourassa2020}. Such approaches become insufficient in the strong-coupling regime, where the interplay between hyperfine interactions and control fields not only imposes additional constraints but also enables faster gate operations and enhanced quantum-node functionality. Nevertheless, systematic demonstrations of strong-coupling-enabled, universal electron–nuclear control in SiC are still lacking. Furthermore, decoherence-protected control techniques have so far been predominantly demonstrated in deeply embedded NV centers that are relatively well isolated from surface-related noise, leaving their applicability to shallow, ion-implanted defects in SiC—crucial for integration with nanophotonic cavities and waveguides~\cite{Crook2020,Calusine2016,Momenzadeh2020Nano,Shahbazi2020APL,Tyler2025APL}—largely unexplored.

In this work, we demonstrate a fully functional quantum node in SiC, based on a hybrid system of electron and strongly coupled nuclear spins. The electron spin of a shallow PL6 color center, created through ion implantation, serves as the processor qubit, while a nearby strongly coupled silicon nuclear spin functions as the memory qubit. Specifically, we design a composite pulse sequence that combines dynamical decoupling and radiofrequency control (DDRF), enabling universal gate operations between the processor and memory while preserving the coherence of the hybrid system. Using this approach, we deterministically prepare and manipulate entangled states on the quantum node with a fidelity of up to $90 \pm 3\%$, exceeding the fault-tolerance threshold required for some quantum network architectures~\cite{Nickerson2013,Nickerson2014}.

\begin{figure}[htbp]
\centering
\includegraphics[scale = 0.5]{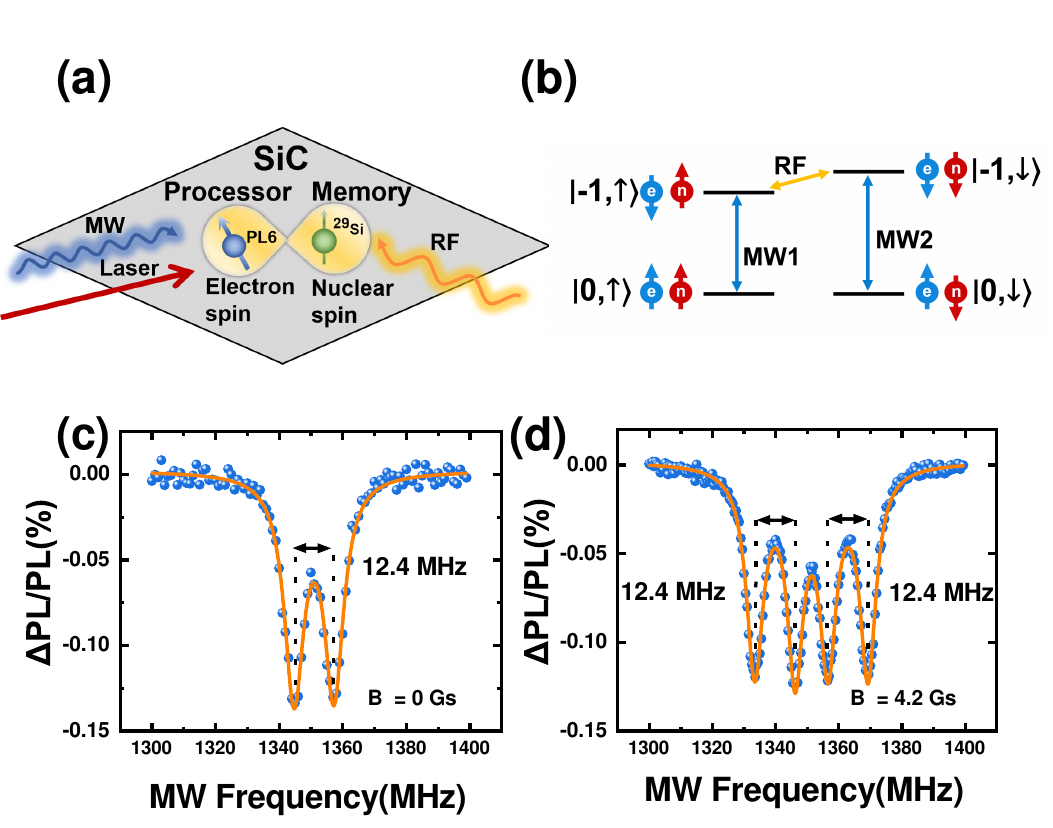}
  \caption{Energy levels of a PL6 center strongly coupled with a $^{29}$Si nuclear spin in a silicon carbide quantum node. (a) Schematic illustration of nuclear spin-memory qubit surrounding a PL6 defect-processor qubit. MW: microwave; RF: radiofrequency pulses. (b) Simplified energy-level diagram of the hybrid electron–nuclear spin register. MW1 and MW2 drive nuclear spin-conserving electron spin transitions, while the RF pulse drives an electron spin-conserving nuclear spin transition. (c) and (d) Optically detected magnetic resonance (ODMR) spectra recorded at zero magnetic field ($B = 0$ Gs) and under a small external field ($B = 4.2$ Gs) aligned with the PL6 symmetry axis. The splitting of 12.4\,MHz shown in the figure represents the hyperfine coupling constant of the nuclear spin. Blue dots indicate raw data; orange lines are Lorentzian fits.} 
\label{Figure 1}
\end{figure}
  
\vspace{10pt}\noindent\textsf{\textbf{\large 2. Experimental implementation and result analysis }}

{ \it 2.1. Quantum nodes based on 4H-SiC PL6 defects}

{Our SiC quantum nodes consist of two fundamental components: a PL6 color center with fast control serving as the processor qubit, and an adjacent nuclear spin with a long lifetime functioning as the memory qubit (Fig.~\ref{Figure 1}). This hybrid system benefits from long coherence times and strong coupling, offering advantages for fast quantum gates and error suppression. The PL6 color center is a distinct type of spin defect with properties resembling those of conventional divacancies in SiC~\cite{Koehl2011}, which have attracted significant interest. Although its exact structure remains under debate~\cite{Son2022}, the PL6 center demonstrates a high optically detected magnetic resonance (ODMR) contrast of 30\% and a photon count rate of 150 kcps at room temperature \cite{Li2022}. It hosts electronic states characterized by near-telecom-wavelength optical transitions and supports gigahertz-frequency spin control, making it a promising candidate comparable to the well-known diamond nitrogen-vacancy center.

Our SiC sample is of the 4H polytype and has a natural isotopic composition, with approximately 1.1\% of carbon atoms ($^{13}\mathrm{C}$) and 4.7\% of silicon atoms ($^{29}\mathrm{Si}$) possessing nuclear spins ($I=1/2$), which couple to the electron spin of the PL6 center via hyperfine interaction (see Supplementary Note 1 in the Supplemental Materials)\cite{SupplementalMaterial}. The PL6 defect features a spin-triplet ground state ($S=1$), where the electron spin can be optically initialized and read out using laser pulses, and coherently manipulated by microwave pulses. The ground-state Hamiltonian of this coupled spin system is given by:
\begin{equation*}
\mathbf{H}=\gamma_e \mathbf{B} \cdot \mathbf{S} + \gamma_n \mathbf{B} \cdot \mathbf{I} + D_{GS} S_Z^2 + \mathbf{S} \cdot \mathbf{A} \cdot \mathbf{I},
\end{equation*}
where $\mathbf{S}$ and $\mathbf{I}$ are the spin-1 operator for the electron and spin-$\frac{1}{2}$ operator for the nuclear spins, respectively. The gyromagnetic ratios are $\gamma_e = 2.8$ $\mathrm{GHz/T}$ for the electron spin and $\gamma_n = 8.5$ $\mathrm{MHz/T}$ for the $^{29}\mathrm{Si}$ nuclear spin. $D_{GS}$ is the zero-field splitting parameter, $\mathbf{B}$ is the applied magnetic field, and $\mathbf{A}$ is the hyperfine interaction tensor. The four terms correspond to the electron Zeeman effect, nuclear Zeeman effect, electron zero-field splitting, and electron-nuclear hyperfine coupling, respectively. The electron spin states are labeled by $m_s = {0, +1, -1}$, and the nuclear spin states by $m_I = {\uparrow, \downarrow}$.

Under an applied magnetic field, the $m_s = \pm 1$ states of the electron spin split. We select the electron spin subspace of $m_s = 0$ and $m_s = -1$ along with the $^{29}\mathrm{Si}$ nuclear spin states $m_I = \uparrow$ and $m_I = \downarrow$, thereby forming a four-level hybrid processor-memory system (Fig.~\ref{Figure 1}b). Because the hyperfine interaction strength decays rapidly with the distance between the electron and nuclear spins, distinct nuclear spins can be spectrally resolved in ODMR by their unique hyperfine splittings. In general, strongly coupled $^{29}\mathrm{Si}$ nuclear spins near the spin defect can be categorized into symmetry-equivalent lattice positions, such as $\rm{Si_{Ia}, Si_{IIa}, Si_{IIb} }$~\cite{Falk2015} (see Supplementary Note 2 in the Supplemental Materials).

We employ a plasmon-enhanced experimental setup to carry out the measurements~\cite{Zhou2023} (see Supplementary Note 1 in the Supplemental Materials)\cite{SupplementalMaterial}. Fig.~\ref{Figure 1}c shows the zero-field ODMR spectrum of the PL6 center coupled to a $\rm{Si_{IIa}}$ nuclear spin, revealing a hyperfine splitting of 12.4 MHz. Fig.~\ref{Figure 1}d displays the ODMR spectrum under a small magnetic field ($B = 4.2$ $\mathrm{Gs}$), where four distinct peaks emerge due to the Zeeman splitting of the $m_s = \pm 1$ states. This confirms that the observed spectral features originate from hyperfine coupling with nuclear spins rather than from strain-induced effects.

{\vspace{10pt}\noindent\textsf{\textbf{\large }\it 2.2. Coherent control of individual devices in quantum nodes}

Quantum nodes require two basic functions: processors and memory that can be
initialized and manipulated separately. More importantly, processors and
memory must be able to efficiently transfer states and manipulate
entanglement.
First, we utilize electron-assisted dynamic nuclear polarization (DNP) to initialize the nuclear spins of the quantum memory~\cite{Smeltzer2009RobustControl,Jacques2009DynamicPolarization,Fischer2013OpticalPolarization,Fischer2013BulkPolarization}. In the process of DNP, the electron spin state can be transferred to the nuclear spin via hyperfine interactions (see Supplementary Note 2 in the Supplemental Materials)\cite{SupplementalMaterial}. When the system is at the excited state
level anticrossing (ESLAC) condition ($B=330\,\text{Gs}$), the hyperfine
interaction induces hybridization between the $|0\downarrow \rangle $ and $%
|-1\uparrow \rangle $ states. This hybridization allows the $|0\downarrow
\rangle $ state to evolve into the $|-1\uparrow \rangle $ state during each
optical cycle. Due to the conservation of nuclear spin ($%
m_{I}=\uparrow $), continuous optical pumping initializes the electron spin
into the $|0\rangle $ state. Therefore, by applying a laser pulse lasting
only a few microseconds, simultaneous polarization of both the electron and
nuclear spins into the $|0\uparrow \rangle $ initial state can be achieved.

Fig.~\ref{Figure 2}a presents the ODMR spectrum of a PL6 center coupled with nuclear
spins, measured at the ESLAC condition ($B = 330 \, \text{Gs}$). We quantify
the nuclear spin polarization using the polarization rate, $P=(N_{\uparrow }-N_{\downarrow })/(N_{\uparrow }+N_{\downarrow })$,
where $N_{\uparrow }$ and $N_{\downarrow }$ represent the populations of
nuclear spins in the $\uparrow $ and $\downarrow $ states, respectively. We
achieve a maximum nuclear spin polarization rate of $99\pm 1\%$ for the
quantum memory (see Supplementary Note 2 in the Supplemental Materials)\cite{SupplementalMaterial}.

Using the polarized nuclear spins, we perform optically detected nuclear
magnetic resonance (ODNMR) experiments to further investigate the coupled
electron-nuclear spin system in PL6. Since the time-averaged
photoluminescence (PL) intensity varies depending on the nuclear spin state,
we can probe the energy difference between the $|-1\uparrow \rangle $ and $%
|-1\downarrow \rangle $ states by designing specific microwave and
radiofrequency (RF) sequences.

To begin, we initialize the system to the $|0\uparrow \rangle $ state using
a 50~$\mu $W laser pulse. A selective microwave (MW1) $\pi $ pulse is then
applied to drive the electron spin transition to the $|-1\uparrow \rangle $
state. Subsequently, a RF pulse with varying frequency is applied to probe
the nuclear spin transition from $\uparrow $ to $\downarrow $. Finally,
another MW1 $\pi $ pulse transfers the system back from $|-1\uparrow \rangle $
to $|0\uparrow \rangle $, followed by optical readout. The resulting ODNMR spectrum (Fig.~\ref{Figure 2}b) reveals resonances at 12.145 MHz, corresponding to hyperfine-coupled nuclear spins located at $\rm{Si_{IIa}}$. 

The distinctive hyperfine coupling and robust spin polarization enable
efficient manipulation and coherent control of individual nuclear spins of
the quantum memory. We further measure the Rabi oscillations of the nuclear
spins. Instead of varying the RF frequency, we set it to resonance and
measure the nuclear spin response by extending the RF pulse duration.
Fig.~\ref{Figure 2}c presents the Rabi oscillations of the nuclear spin in the $|-1\rangle $ electron spin state, with the fitted $\pi$-pulse duration
determined to be 2.328~$\mu$s. Additionally, we perform Ramsey
interferometry measurements on the nuclear spin (Fig.~\ref{Figure 2}d). By fitting the
Ramsey oscillations, we estimate the inhomogeneous coherence time ($\rm{T_{2n}^{\ast }}$) of the nuclear spins at room temperature to be
approximately 143~$\mu $s, nearly seventy times longer than that of the electron spins ($\rm{T_{2e}^{\ast
}}=2.04\,\mu $s, see Supplementary Note 2 in the Supplemental Materials)\cite{SupplementalMaterial}.

It is important to note that the high polarization of the nuclear spins
makes it challenging to precisely extract the MW2 transition frequency of the quantum processor from the ODMR spectrum. To address this, we first
polarize the system and sequentially apply a MW1 $\pi $ pulse and a RF $\pi $
pulse, thereby preparing the system in the $|-1\downarrow \rangle $ state. By
Scanning the MW frequency, we can accurately determine the MW2 transition
frequency ($|-1\downarrow \rangle \rightarrow |0\downarrow \rangle $) and the
corresponding MW2 $\pi $-pulse parameters (see Supplementary Note 2 in the Supplemental Materials)\cite{SupplementalMaterial}.

\begin{figure}[htbp]
\centering
\includegraphics[scale = 0.5]{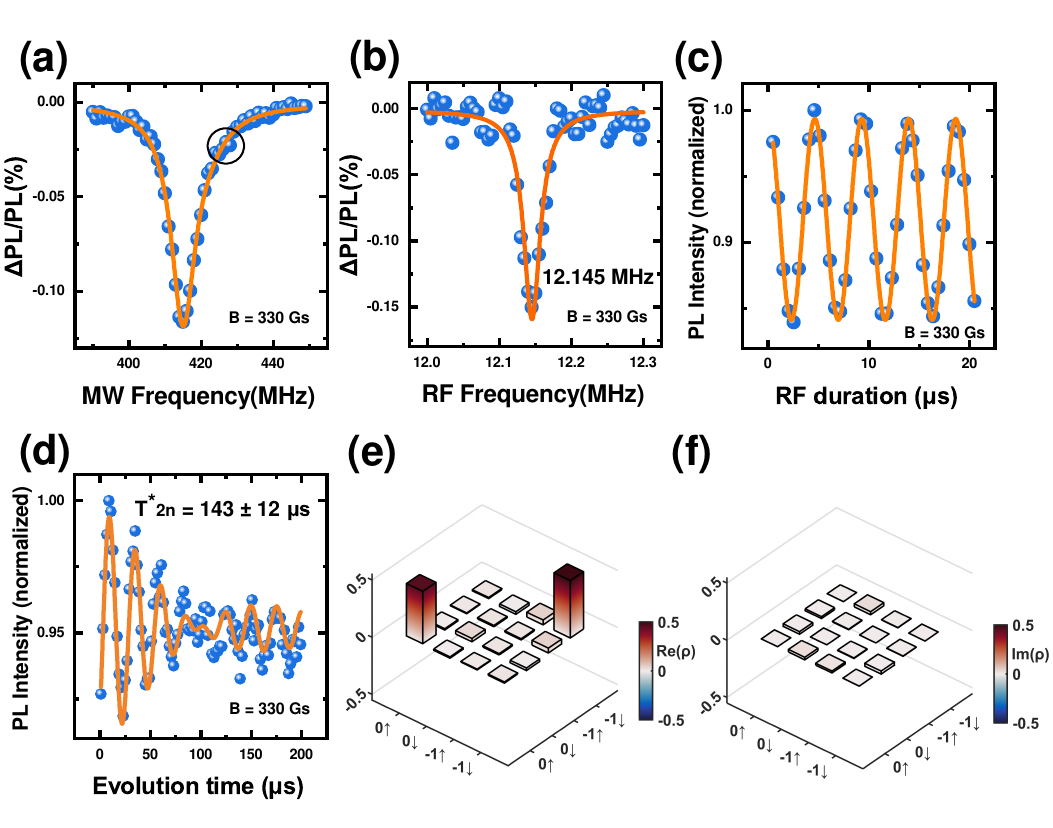}
  \caption{Coherent Control of Nuclear Spins and Quantum State Tomography of Hybrid Entangled States. (a) The ODMR spectra recorded at a 330 Gs magnetic field applied along the direction of the PL6
defect. The solid-line circle highlights the disappearance of the right branch of the hyperfine-split resonance. (b) The ODNMR spectrum. A peak at around 12.145 MHz is observed, indicating a \(\rm{Si_{IIa}}\) nuclear spin coupled with a single PL6. (c) Nuclear Rabi oscillations. (d) The Nuclear Ramsey fringes are fitted with a double cosine exponential decay function. Blue dots represent raw data, and orange lines indicate the Lorentzian fits. (e) and (f) The real and imaginary parts of the entangled-state density matrix reconstructed via maximum likelihood estimation are shown. The total preparation time is 14.7\,$\mu$s, and the resulting fidelity is only $70 \pm 3\%$.
} 
 
 \label{Figure 2}
\end{figure} 

Now we focus on the entanglement manipulation between the processor and the
memory. Having precisely characterized the MW1, MW2, and RF transitions, we implement the controlled-not gates (CNOT) within the hybrid system: \(\text{C}_n \text{NOT}_e\) and \(\text{C}_e \text{NOT}_n\), where the control and target qubits are the nuclear and electron spins, respectively. This is a critical step toward Bell state preparation between the processor and the memory. We first initialize the system and apply an MW1 \(\pi/2\) pulse to preparing: 
\(\frac{1}{\sqrt{2}}(\ket{0,\uparrow}+\ket{-1,\uparrow}).\)     
A subsequent RF $\pi$-pulse flips the nuclear spin conditionally, yielding: 
\(\frac{1}{\sqrt{2}}(\ket{0,\uparrow}+\ket{-1,\downarrow})\). We characterize the resulting Bell state using quantum state tomography at room temperature, without the need for cryogenic cooling. The statistical uncertainties of the reconstructed density matrices, as well as the derived fidelity, are estimated using a Monte Carlo method that accounts for Poissonian photon-counting noise in the tomographic measurements. Details of the error bar estimation procedure are provided in the Supplementary Material. Because of the short electron spin coherence time, the fidelity of the Bell state is highly sensitive to the total gate duration (see Supplementary Note 4 and 5 in the Supplemental Materials)\cite{SupplementalMaterial} . The state fidelity is calculated as $F=\sqrt{\sqrt{\rho_{e}}\rho_{t}\sqrt{\rho_e}}$, where $\rho_{e}$ and $\rho_{t}$ denote the experimentally recontructed and theoretically ideal density matrices, respectively. In the experiment, the total preparation time is set to 14.7 $\mu$s, which significantly exceeds the electron spin coherence time. Fig.~\ref{Figure 2}e and Fig.~\ref{Figure 2}f present the real and imaginary parts of the density matrix of the prepared  Bell state, with the measured fidelity reaching approximately $70 \pm 3\%$.

{\it 2.3. Decoherence-protected gate operations between processor and memory}

Typical gate operations can only generate low-fidelity entangled states, highlighting the necessity of developing decoherence-protected gate operations for this hybrid device. The electron spin of the processor and the nuclear spin of the memory exhibit vastly different control times and coherence times, posing new challenges for the precise manipulation of this hybrid quantum node. In particular, the dephasing time $\rm{T_{2e}^{\ast }}$ of the PL6 electron spin is inherently short, while nuclear spin manipulations generally require operation times that exceed the electron spin dephasing time.

\begin{figure}[htbp]
\centering
\includegraphics[scale = 0.5]{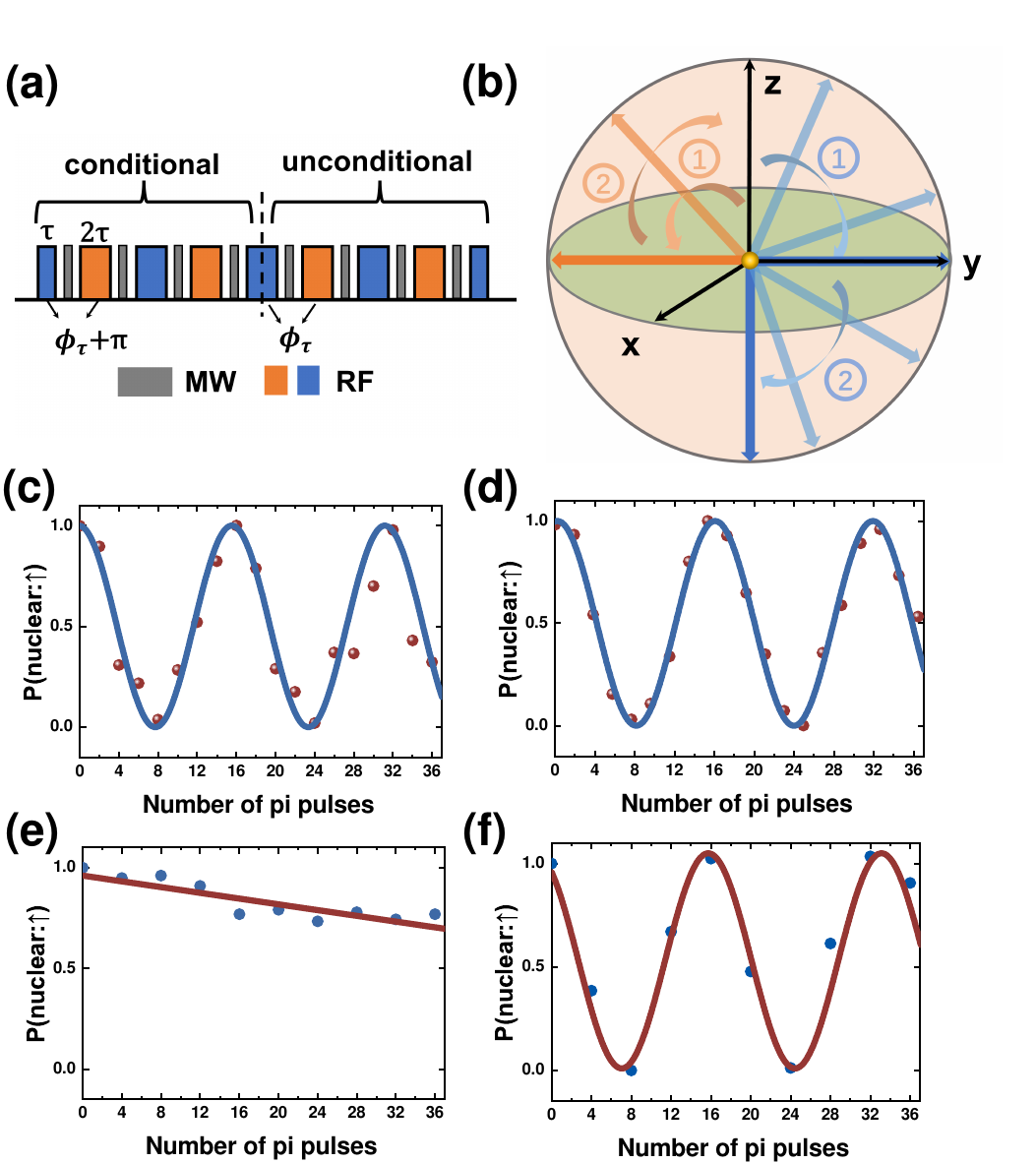}
\caption{Implementing decoherence-protected quantum gates for electron-nuclear spin registers using DDRF. (a) Schematic of the DDRF gate pulse sequence. Microwave (MW) pulses for the electron spin (gray) are interleaved with radiofrequency (RF) pulses (blue and orange, denoting odd and even RF numbers, respectively), enabling selective control of a single nuclear spin. The parameters $\tau$ and $\phi_\tau = A_{zz}\tau$ represent the RF pulse duration and the applied phase correction, respectively.
(b) Rotation of the nuclear spin under the pulse sequence in (a). The initial nuclear spin state (coincide with the z-axis) undergoes either a complete $\pi$ rotation (blue arrow) or remains unchanged (orange arrow) depending on whether the electron spin is initially in state $|-1\rangle$ or $|0\rangle$, respectively. Circles 1 and 2 correspond to the applied conditional and unconditional pulses illustrated in (a). (c)-(d) Unconditional nuclear gate calibrations when the electron spin is initialized in either the \( \left| 0 \right\rangle \) (c) or \( \left| -1 \right\rangle \) (d) state, the nuclear spin undergoes identical rotations. (e)-(f) Conditional nuclear gate calibration. When the electron spin is initialized in the \( \left| 0 \right\rangle \) state, the nuclear spin exhibits no rotation around the z-axis (e). However, when the electron spin is initialized in the \( \left| -1 \right\rangle \) state, the nuclear spin undergoes a controlled rotation around the effective $x$-axis, with the rotation angle increasing as the number of pulses \( N \) increases (f).
}
\label{Figure 3}
\end{figure} 

We implement a composite pulse sequence that combines dynamic decoupling (DD) microwave pulses~\cite{Ryan2010,deLange2010,Taminiau2012,Cramer2016,Maity2022,Nguyen2019} and radiofrequency (RF) pulses~\cite{Bradley2019,vanDerSar2012,Beukers2025}. The RF frequency is tuned to the nuclear spin transition corresponding to the electron spin state $|-1\rangle$. The sequence is constructed by inserting RF pulses into each idle time $\tau $ of the DD sequence, forming the pattern $(\tau
-\pi -2\tau -\pi -\tau )^{\frac{N}{4}}$, where $N$ is the number of gates. 

This pulse sequence enables the realization of two fundamental gate operations between the processor and memory qubits. To implement the controlled nuclear spin rotation gate, $\text{C}_{e}\text{ROT}_{n}$,
the parameter $\tau $ must satisfy specific constraints, $\tau =\frac{(2n+1)\pi }{A}$, where $n$ is an integer and $A$ is the hyperfine splitting constant. This condition ensures that the rotation axis of the nuclear spins on the Bloch sphere are antiparallel under the two electron spin states ($|0\rangle$ and $|-1\rangle $), thereby achieving the conditional rotation gate $\text{C}_{e}\text{ROT}_{n}(\pm\theta )$.

The physical origin of this gate lies in the time difference $\tau$ of the effective RF-driven manipulation of the nuclear spin under different electron spin states. During this time interval, the nuclear spin precessions accumulate a relative phase difference of $\pi $, resulting in rotations along diametrically opposed axes on the Bloch sphere-this is the essence of conditional nuclear spin rotation. Conversely, when the timing satisfies $\tau=\frac{2n\pi }{A}$, the nuclear spin rotation axes under both electron spin states become parallel, realizing an unconditional nuclear rotation gate, denoted as $\text{ROT}_{n}(\theta )$.

While this approach demonstrates theoretical viability, it imposes stringent requirements on temporal precision. The parameter $\tau$ must be precisely controlled, with the required accuracy scaling proportionally to the hyperfine coupling strength $A$. In the previous study \cite{vanDerSar2012}, the value of $A$ was approximately 2 MHz, which allowed for a more relaxed timing design. However, in our system, $A = 12.4$ MHz, which is roughly six times the strength of previous studies, where the characteristic time $2\pi/A \approx 80$ ns, achieving acceptable time error necessitates sub-0.5 ns timing resolution. Furthermore, the constrained $\tau = \frac{(2n+1)\pi}{A}$ relationship fundamentally limits the selection of the total operation time, introducing practical implementation challenges.

To overcome this limitation, we implement a phase-compensated RF protocol by adjusting the phases of the RF pulses, where \( \phi _{\tau} \) denotes the compensatory phase applied in the modified RF pulse sequence.As illustrated in Fig.~\ref{Figure 3}a, the electron spin undergoes dynamical decoupling via microwave (MW) \( \pi \)-pulses (gray), interleaved with RF pulses (blue and orange), which correspond to alternating RF segments with incremented phases. This interleaved design enables selective and phase-coherent control of a single nuclear spin conditioned on the electron state. The initial electron spin state determines which RF pulses are resonant with the nuclear spin. Specifically, when the electron spin is in \( | -1 \rangle \), the nuclear spin is rotated by the odd-numbered RF pulses (blue), whereas for the electron spin in \( | 0 \rangle \), the even-numbered RF pulses (orange) become resonant. To avoid unwanted detuning, RF pulses are switched off during the electron spin \( \pi \)-pulses. By fine-tuning the RF phases \( \phi_\tau \), both conditional and unconditional gates can be implemented without requiring precise control of the inter-pulse delay \( \tau \) (see Supplementary Note 7 in the Supplemental Materials) \cite{SupplementalMaterial}.

Fig.~\ref{Figure 3}b demonstrates an example of nuclear spin rotations under the sequence shown in Fig.~\ref{Figure 3}a. Circles~1 and~2 represent the conditional and unconditional operations, respectively. In the conditional operation (first half of the sequence), the nuclear spin undergoes a \( \pi/2 \)  clockwise rotation around the effective \( x \)-axis when the electron spin is initially in \( | -1 \rangle \) (blue circle~1), and a \( -\pi/2 \) counterclockwise rotation when the electron spin is initially in \( | 0 \rangle \) (orange circle~1). In the unconditional operation (second half of the sequence), the nuclear spin experiences a \( \pi/2 \) clockwise rotation around the \( x \)-axis irrespective of whether the electron spin starts in \( | 0 \rangle \) or \( | -1 \rangle \) (blue and orange circle~2). Combining these two operations yields an electron-controlled nuclear spin flip: when the electron spin is in \( | -1 \rangle \), the nuclear spin undergoes a full \(\pi\) rotation around the \( x \)-axis, whereas for the electron spin in \( | 0 \rangle \), the nuclear spin remains unchanged.

To evaluate the performance of our DDRF pulses, we further investigate the calibration of two basic gate operations of the quantum node: conditional and unconditional nuclear spin rotations. Instead of initializing the electron spin in a $\frac{1}{\sqrt{2}}(\left\vert 0\right\rangle +\left\vert -1\right\rangle)$, we prepare the system in the $\left\vert 0,\uparrow \right\rangle $ and $\left\vert -1,\uparrow \right\rangle $ states separately. After optimizing the timing parameter $\tau$, RF power, and pulse phases, we characterize the spin control dynamics.

We begin by characterizing the unconditional rotation, $\text{ROT}_{n}(\theta)$. For direct comparison with the conditional case discussed below, we apply the unconditional rotation twice to realize $\text{ROT}_{n}(2\theta)$, as shown in Fig.~\ref{Figure 3}c,Fig.~\ref{Figure 3}d. Cosine-like oscillations are observed in the nuclear spin transition probabilities as a function of the number of gate units, independent of the initial electron spin state ($\left\vert 0,\uparrow \right\rangle$ or $\left\vert -1,\uparrow \right\rangle$). These oscillations resemble conventional RF-driven nuclear Rabi oscillations, but exhibit distinct brightness contrasts between the $\left\vert 0,\uparrow \right\rangle/\left\vert 0,\downarrow \right\rangle$ and $\left\vert -1,\uparrow \right\rangle/\left\vert -1,\downarrow \right\rangle$ state populations. This behavior confirms the successful implementation of the unconditional gate operation.

Next, we characterize the conditional rotation (Fig.~\ref{Figure 3}e, Fig.~\ref{Figure 3}f). For conditional $\text{C}_{e}\text{ROT}_{n}(\pm \theta )$ operations, direct fluorescence
measurements yield identical cosinusoidal patterns similar to the unconditional case, due to the optical indistinguishability of rotations by $\pm \theta $. To break this degeneracy, we apply a sequence combining both conditional and unconditional operations:
\begin{equation*}
\text{ROT}_{n}(\theta )\text{C}_{e}\text{ROT}_{n}(\pm \theta )=\text{C}_{e}%
\text{ROT}_{n}(0,2\theta ).
\end{equation*}%
Fluorescence measurements after this sequence reveal a static nuclear spin population when the system is initialized in the $\left\vert 0,\uparrow \right\rangle $ state (indicating no nuclear spin rotation), while a rotation of $2\theta $ is observed for the $\left\vert -1,\uparrow
\right\rangle $ state. This clear differential response demonstrates the successful realization of state-selective nuclear spin control: electron
spin-dependent nuclear manipulation enabled by the DDRF sequence. The observed $2\theta $ oscillations, together with the absence of rotation in $\left\vert 0,\uparrow \right\rangle $ state, provide direct experimental confirmation of the DDRF protocol's capability to achieve conditional quantum gate operations, fulfilling a key requirement for spin-based quantum information processing.

{\it 2.4. High fidelity entanglement control in quantum nodes}

We then investigate the coherence-preserving capability of this pulse sequence for the hybrid quantum node. Standard coherence time measurements typically involve sweeping the inter-pulse delays between microwave $\pi$-pulses. However, in the DDRF scheme, the nuclear spin rotation angles inherently depend on these temporal intervals, making simultaneous RF application incompatible during coherence characterization.
Therefore, we revert to conventional dynamical decoupling (DD) sequences for
electron spin coherence measurements. Specifically, we initialize the system in the $\left\vert 0,\uparrow \right\rangle $ state and apply the microwave
sequence:$\frac{\pi }{2}-\left( \tau -\pi -2\tau -\pi -\tau \right) ^{\frac{N}{2}}-\frac{\pi }{2}$ without applying RF fields. We measure the electron spin dephasing time as a function of the number of applied $\pi$-pulse numbers $N$ (Fig.~\ref{Figure 4}a), using this giant $\pi$-pulse sequence. The results show a progressive extension of the electron spin dephasing times, increasing from $\rm{T_{2}}=8.2\pm0.4\,\mu s$ at $N=1$ to $\rm{T_{2}^{DD}}=31.7\pm2.3\,\mu s$
at $N=16$, demonstrating substantial coherence preservation.

\begin{figure}[htbp]
\centering
\includegraphics[scale = 0.5]{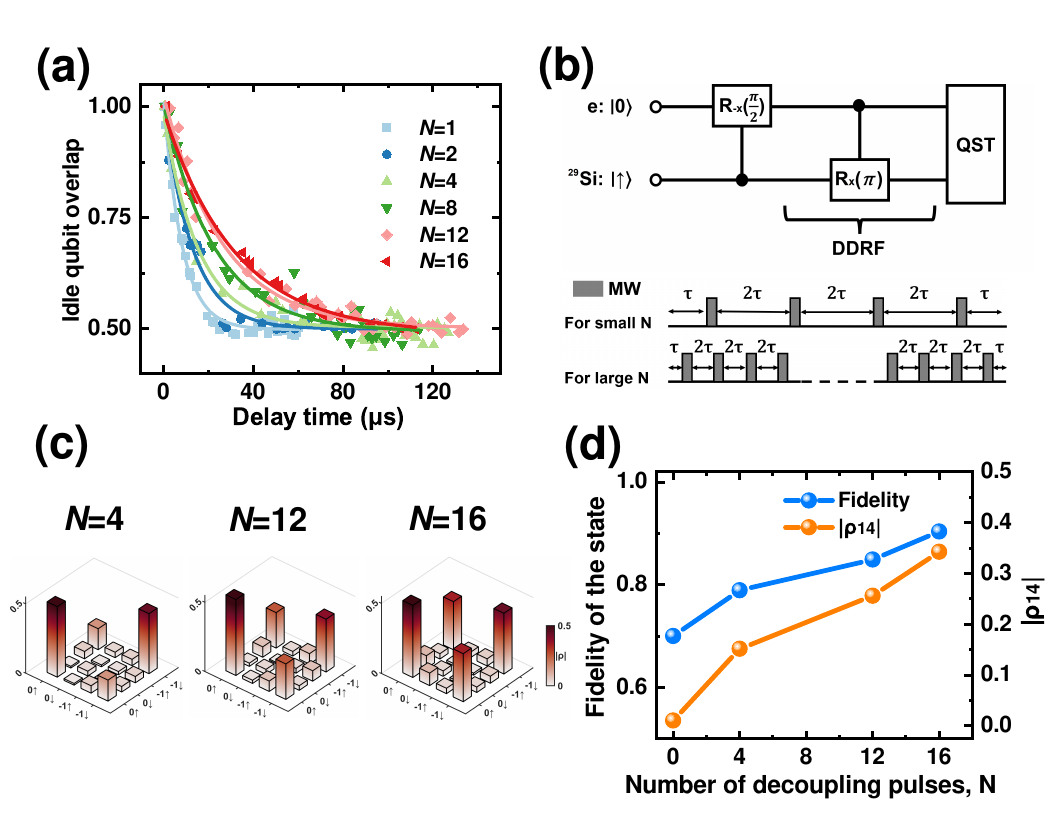}
\caption{The performance of DDRF decoherence-protected gates. (a) The measurement of electron spin coherence is conducted using dynamical decoupling sequences with varying numbers of \( \pi \) pulses \( N \). (b) The entangled state preparation sequence and the sequence to validate the effectiveness of the decoherence-protected gate. (c) The reconstructed absolute density matrices (\( | \rho | \)) after applying the DDRF decoherence-protected gate. The quantum state tomography results for gate numbers \( N = 4, 12, 16 \) are displayed from left to right, demonstrating a significant enhancement in fidelity. (d) The fidelity and the magnitude of the off-diagonal element ($\rho_{14}$) of the Bell state as a function of the number of decoupling pulses \( N \). }
\label{Figure 4}
\end{figure}

Furthermore, we utilize this decoherence-protected gate operation to generate entangled states between the processor and memory. By
implementing the DDRF protocol with optimized timing and phase parameters, and setting $2\theta =\pi $ (corresponding to the $\text{C}_{e}\text{ROT}_{n}$ gate configuration), we prepare the electron-nuclear spin entangled state: $\left\vert \psi \right\rangle =\frac{1}{\sqrt{2}}\left( \left\vert
0,\uparrow \right\rangle +\left\vert -1,\downarrow \right\rangle \right) $ using a $\text{MW1}$ $\pi /2$ pulse followed by an $\text{RF}$ $\pi$ pulse (Fig.~\ref{Figure 4}b). To explicitly demonstrate the coherence-preserving capability of the DDRF protocols under large numbers of microwave pulses, we maintain a fixed total gate
duration of approximately $14.7\,\mu s$-almost twice the Hahn
echo $\rm{T_{2}}$ limit-while increasing the number of $\pi $-pulses from $N=4$ to $N=16$ (see Supplementary Note 6 in the Supplemental Materials) \cite{SupplementalMaterial}. The inter-pulse delay $\tau $ is simultaneously adjusted to keep the total duration constant during entangled state
preparation. After each implementation, we perform quantum state tomography. As shown in Fig.~\ref{Figure 4}c, the reconstructed Bell state density matrices ($|\rho|$) for $N=4$, 12, and 16 are displayed from left to right. The tomography results reveal that as $N$ increases, the magnitude of the off-diagonal
element $\rho_{14}$,  corresponding to the coherence term $\rho(|0\uparrow\rangle\langle-1\downarrow|)$) of the density matrix, also increases. This indicates a gradual
recovery toward the ideal case-a trend clearly visible in Fig.~\ref{Figure 4}d. Using the fidelity calculation formula, we obtain a Bell state fidelity of $90 \pm 3\%$ for $N=16$ (Fig.~\ref{Figure 4}d). Notably, this fidelity exceeds the fault-tolerance threshold required by certain quantum network architectures \cite{Nickerson2013,Nickerson2014}. These results demonstrate that
even when the total state preparation time far surpasses the electron spin dephasing time $\rm{T_{2e}^{\ast }}$, our decoherence-protected gate enables the deterministic, high-fidelity entangled state preparation in hybrid quantum nodes, providing a crucial resource for scalable quantum networks. Compared with typical NV–nuclear spin systems in diamond, the PL6–nuclear spin system in SiC exhibits significantly stronger and anisotropic hyperfine interactions, reaching up to $\sim$12.4~MHz. This regime necessitates precisely calibrated phase compensation and accurate timing synchronization to correctly track the hyperfine-induced nuclear-spin precession. As a result, the control scheme implemented for PL6 centers differs both quantitatively and operationally from previously reported NV-based protocols.

\vspace{10pt}\noindent\textsf{\textbf{\large 3. Discussion and Conclusion}}

In this work, we demonstrate a fully functional quantum node composed of an electron spin processor and a nuclear spin memory in SiC. This hybrid node employs composite pulse sequences that combine tunable radiofrequency pulses and dynamic decoupling pulses, enabling universal gate operations on both the processor and the memory while preserving the coherence of the entire quantum system.  Under realistic experimental conditions, we achieve entangled state preparation with a fidelity reaching 90\%, surpassing the fault-tolerance threshold required by certain quantum network architectures.

Entanglement between electron spins and strongly coupled nuclear spins has been achieved in both single-defect systems~\cite{Bourassa2020,Hu2024} and ensembles~\cite{Klimov2015} in SiC, but the implementation of universal two-qubit gates that are resilient to decoherence has remained elusive. Our work addresses this critical gap by demonstrating entangling gates in a SiC quantum node that maintain coherence through carefully engineered pulse sequences. By leveraging the mature materials platform and established semiconductor processing technologies of SiC, this approach provides a scalable and technologically accessible pathway toward robust quantum information processing. Combined with optical interfaces, the fully functional quantum node realized here may serve as a fundamental building block for near-term quantum network applications. Looking ahead, further improvements are expected from isotopic engineering of the host crystal, which would suppress the nuclear spin bath, extend the electron-spin coherence time, and thereby mitigate high-frequency noise contributions that currently limit gate performance.

\vspace{10pt}\noindent\textsf{\textbf{\large Data availability statement}}

All data that support the findings of this study are included  within the article (and any supplementary files).

\vspace{10pt}\noindent\textsf{\textbf{\large Acknowledgements}}

This work was supported by the Innovation Program for Quantum Science and Technology (No.\ 2021ZD0301400 and No.\ 2021ZD0301200), National Natural Science Foundation of China (No.\ 92365205, No. W2411001, and No.\ 12350006), USTC Major Frontier Research Program (No.\ LS2030000002) and USTC (NO.\ YD2030002026). This work was partially performed at the University of Science and Technology of China Center for Micro and Nanoscale Research and Fabrication.  

\vspace{10pt}\noindent\textsf{\textbf{\large Author contributions}}

J.-S.X and C.-F.L conceived the experiments. S.R. and R.-J.L built the experimental setup and performed the measurements with the help of Z.-X.H., J.-Y.Z., W.-X.L, Z.-H.H and B.C. T.T. provided the theoretical analysis and support. S.R., R.-J.L., J.-S.X. and T.T. wrote the paper with contributions from all co-authors. J.-S.X. C.-F.L. and G.-C.G. supervised the project. All authors contributed to the discussion of the results.  

\vspace{10pt}\noindent\textsf{\textbf{\large Competing interests}}

The authors declare no competing interests.


\begin{thebibliography}{50}

\bibitem{Kimble2008}
H. J. Kimble. The quantum internet. \emph{Nature} \textbf{453}, 1023–1030 (2008).

\bibitem{Sangouard2011}
N. Sangouard \emph{et al.} Quantum repeaters based on atomic ensembles and linear optics. \emph{Rev. Mod. Phys.} \textbf{83}, 33–80 (2011).

\bibitem{Northup2014}
T. E. Northup \& R. Blatt. Quantum information transfer using photons. \emph{Nat. Photon.} \textbf{8}, 356–363 (2014).

\bibitem{Wehner2018}
S. Wehner \emph{et al.} Quantum internet: A vision for the road ahead. \emph{Science} \textbf{362}, eaam9288 (2018).

\bibitem{Awschalom2018}
D. D. Awschalom \emph{et al.} Quantum technologies with optically interfaced solid-state spins. \emph{Nat. Photonics} \textbf{12}, 516–527 (2018).

\bibitem{Rodgers2018}
L. V. H. Rodgers \emph{et al.} Materials challenges for quantum technologies based on color centers in diamond. \emph{Nat. Rev. Mater.} \textbf{3}, 38–51 (2018).

\bibitem{Aharonovich2016}
I. Aharonovich \emph{et al.} Solid-state single-photon emitters. \emph{Nat. Photonics} \textbf{10}, (2016).

\bibitem{Reiserer2015}
A. Reiserer \& G. Rempe. Cavity-based quantum networks with single atoms and optical photons. \emph{Rev. Mod. Phys.} \textbf{87}, 1379 (2015).

\bibitem{Duan2010}
L.-M. Duan \& C. Monroe. Colloquium: Quantum networks with trapped ions. \emph{Rev. Mod. Phys.} \textbf{82}, 1209 (2010).

\bibitem{Koehl2011}
W. F. Koehl \emph{et al.} Room temperature coherent control of defect spin qubits in silicon carbide. \emph{Nature} \textbf{479}, 84 (2011).

\bibitem{Christle2015}
D. J. Christle \emph{et al.} Isolated electron spins in silicon carbide with millisecond coherence times. \emph{Nat. Mater.} \textbf{14}, 160–163 (2015).

\bibitem{Seo2016}
H. Seo \emph{et al.} Quantum decoherence dynamics of divacancy spins in silicon carbide. \emph{Nat. Commun.} \textbf{7}, 12935 (2016).

\bibitem{Christle2017}
D. J. Christle \emph{et al.} Isolated spin qubits in SiC with a high-fidelity infrared spin-to-photon interface. \emph{Phys. Rev. X} \textbf{7}, 021046 (2017).

\bibitem{Anderson2019}
C. P. Anderson \emph{et al.} Electrical and optical control of single spins integrated in scalable semiconductor devices. \emph{Science} \textbf{366}, 1225–1230 (2019).

\bibitem{Xu2015SiCnetwork}
J.-S. Xu \emph{et al.} Silicon carbide based quantum networking. \emph{Science Bulletin} \textbf{60}, 48--55 (2015).

\bibitem{Crook2020}
A. L. Crook \emph{et al.} Purcell enhancement of a single silicon carbide color center with coherent spin control. \emph{Nano Lett.} \textbf{20}, 3427–3434 (2020).

\bibitem{Widmann2015}
M. Widmann \emph{et al.} Coherent control of single spins in silicon carbide at room temperature. \emph{Nat. Mater.} \textbf{14}, (2015).

\bibitem{Nagy2019}
R. Nagy \emph{et al.} High fidelity spin and optical control of single silicon vacancy centres in silicon carbide. \emph{Nat. Commun.} \textbf{10}, 1954 (2019).

\bibitem{Babin2022}
C. Babin \emph{et al.} Fabrication and nanophotonic waveguide integration of silicon carbide colour centres with preserved spin-optical coherence. \emph{Nat. Mater.} \textbf{21}, 67–73 (2022).

\bibitem{Wang2020}
J.-F. Wang \emph{et al.} Coherent control of nitrogen-vacancy center spins in silicon carbide at room temperature. \emph{Phys. Rev. Lett.} \textbf{124}, 223601 (2020).

\bibitem{Mu2020}
Z. Mu \emph{et al.} Coherent manipulation with resonant excitation and single emitter creation of nitrogen vacancy centers in 4H silicon carbide. \emph{Nano Lett.} \textbf{20}, 6142–6147 (2020).

\bibitem{Li2022}
Q. Li \emph{et al.} Room-temperature coherent manipulation of single-spin qubits in silicon carbide with a high readout contrast. \emph{Nat. Sci. Rev.} \textbf{9}, nwab122 (2022).

\bibitem{Zhou2023}
J.-Y. Zhou \emph{et al.} Plasmonic-Enhanced Bright Single Spin Defects in Silicon Carbide Membranes. \emph{Nano Lett.} \textbf{23}, 4334–4343 (2023).

\bibitem{Son2022}
N. T. Son, D. Shafizadeh, T. Ohshima, I. G. Ivanov, Modified divacancies in 4H-SiC. \emph{J. Appl. Phys.} \textbf{132}, 025703 (2022).

\bibitem{Falk2015}
A. L. Falk \emph{et al.} Optical polarization of nuclear spins in silicon carbide. \emph{Phys. Rev. Lett.} \textbf{114}, 247603 (2015).

\bibitem{Klimov2015}
P. V. Klimov \emph{et al.} Quantum entanglement at ambient conditions in a macroscopic solid-state spin ensemble. \emph{Sci. Adv.} \textbf{1}, e1501015 (2015).

\bibitem{Bourassa2020}
A. Bourassa \emph{et al.} Entanglement and control of single nuclear spins in isotopically engineered silicon carbide. \emph{Nat. Mater.} \textbf{19}, 1319 (2020).

\bibitem{Fang2024}
R.-Z. Fang \emph{et al.} Experimental generation of spin-photon entanglement in silicon carbide. \emph{Phys. Rev. Lett.} \textbf{132}, 160801 (2024).

\bibitem{Lai2024}
X.-Y. Lai \emph{et al.} Single-shot readout of a nuclear spin in silicon carbide. \emph{Phys. Rev. Lett.} \textbf{132}, 180803 (2024).

\bibitem{Hesselmeier2024siv}
E. Hesselmeier \emph{et al.} Qudit-based spectroscopy for measurement and control of nuclear-spin qubits in silicon carbide. \emph{Phys. Rev. Lett.} \textbf{132}, 090601 (2024).

\bibitem{Hesselmeier2024}
E. Hesselmeier \emph{et al.} High-fidelity optical readout of a nuclear-spin qubit in silicon carbide. \emph{Phys. Rev. Lett.} \textbf{132}, 180804 (2024).

\bibitem{Hu2024}
H. Hu \emph{et al.} Room-temperature waveguide integrated quantum register in a semiconductor photonic platform. \emph{Nat. Commun.} \textbf{15}, 10256 (2024).

\bibitem{Calusine2016}
G. Calusine \emph{et al.} Cavity-enhanced measurements of defect spins in silicon carbide. \emph{Phys. Rev. Appl.} \textbf{6}, 014019 (2016).

\bibitem{Momenzadeh2020Nano}
S. A. Momenzadeh \emph{et al.} Nanoengineered diamond waveguide as a robust bright platform for nanomagnetometry using shallow nitrogen vacancy centers. \emph{Nano Lett.} \textbf{15}, 165--169 (2020).

\bibitem{Shahbazi2020APL}
S. Shahbazi \emph{et al.} Vector magnetometry using shallow implanted NV centers in diamond with waveguide-assisted dipole excitation and readout. \emph{APL Photonics} \textbf{5}, 021301 (2020).

\bibitem{Tyler2025APL}
S. Tyler \emph{et al.} Extended $T_2$ times of shallow implanted NV centers in chemically mechanically polished diamond. \emph{Appl. Phys. Lett.} \textbf{126}, 054001 (2025).

\bibitem{Smeltzer2009RobustControl}
B. Smeltzer \emph{et al.} Robust control of individual nuclear spins in diamond. \emph{Phys. Rev. A} \textbf{80}, 050302 (2009).

\bibitem{deLange2010}
G. de Lange \emph{et al.} Universal dynamical decoupling of a single solid-state spin from a spin bath. \emph{Science} \textbf{330}, 60–63 (2010).

\bibitem{Ryan2010}
C. Ryan, J. Hodges, D. Cory Robust Decoupling Techniques to Extend Quantum Coherence in Diamond. \emph{Phys. Rev. Lett.} \textbf{105}, 200402 (2010).

\bibitem{Taminiau2012}
T. H. Taminiau \emph{et al.} Detection and control of individual nuclear spins using a weakly coupled electron spin. \emph{Phys. Rev. Lett.} \textbf{109}, 137602 (2012).

\bibitem{vanDerSar2012}
T. van der Sar \emph{et al.} Decoherence-protected quantum gates for a hybrid solid-state spin register. \emph{Nature} \textbf{484}, 82–86 (2012).

\bibitem{Bradley2019}
C. E. Bradley \emph{et al.} A ten-qubit solid-state spin register with quantum memory up to one minute. \emph{Phys. Rev. X} \textbf{9}, 031045 (2019).

\bibitem{Nickerson2013}
N. H. Nickerson \emph{et al.} Topological quantum computing with a very noisy network and local error rates approaching one percent. \emph{Nat. Commun.} \textbf{4}, 1756 (2013).

\bibitem{Nickerson2014}
N. H. Nickerson \emph{et al.} Freely scalable quantum technologies using cells of 5-to-50 qubits with very lossy and noisy photonic links. \emph{Phys. Rev. X} \textbf{4}, 041041 (2014).

\bibitem{Jacques2009DynamicPolarization}
V. Jacques \emph{et al.} Dynamic polarization of single nuclear spins by optical pumping of nitrogen-vacancy color centers in diamond at room temperature. \emph{Phys. Rev. Lett.} \textbf{102}, 057403 (2009).

\bibitem{Fischer2013OpticalPolarization}
R. Fischer \emph{et al.} Optical polarization of nuclear ensembles in diamond. \emph{Phys. Rev. B} \textbf{87}, 125207 (2013).

\bibitem{Fischer2013BulkPolarization}
R. Fischer \emph{et al.} Bulk nuclear polarization enhanced at room temperature by optical pumping. \emph{Phys. Rev. Lett.} \textbf{111}, 057601 (2013).


\bibitem{Cramer2016}
J. Cramer \emph{et al.} Repeated quantum error correction on a continuously encoded qubit by real-time feedback. \emph{Nat. Commun.} \textbf{7}, 11526 (2016).

\bibitem{Maity2022}
S. Maity \emph{et al.} Mechanical control of a single nuclear spin. \emph{Phys. Rev. X} \textbf{12}, 011056 (2022).

\bibitem{Nguyen2019}
C. T. Nguyen \emph{et al.} Quantum network nodes based on diamond qubits with an efficient nanophotonic interface. \emph{Phys. Rev. Lett.} \textbf{123}, 183602 (2019).

\bibitem{Beukers2025}
H. K. C. Beukers \emph{et al.} Control of solid-state nuclear spin qubits using an electron spin-1/2. \emph{Phys. Rev. X} \textbf{15}, 021011 (2025).

\bibitem{SupplementalMaterial}
See Supplemental Material for more details.

\end{thebibliography}
\end{document}